\documentclass[showpacs,superscriptaddress, preprintnumbers,amsmath,amssymb]{revtex4}

\usepackage{graphicx}
\usepackage{dcolumn}
\usepackage{bm}
\include{epsf}

\begin{document}

\preprint{SOFIA/COST/3}


 \title{ A scientometrics law about co-authors and their ranking.\\The co-author core.}


\author{M. Ausloos}\email{marcel.ausloos@ulg.ac.be}
\affiliation{Beauvallon, r. Belle Jardiniere, 483, \\
B-4031 Liege, Wallonia-Brussels Federation\\
{\it previously at}\\ 
GRAPES@SUPRATECS, ULG, B5a Sart-Tilman, B-4000 Li\`ege, Euroland
\protect\\ e-mail address: marcel.ausloos@ulg.ac.be }


\date{today}

 \begin{abstract}
Rather than "measuring" a scientist   impact through the number of citations which his/her published work can have generated, isn't it more appropriate to consider his/her value through his/her scientific network performance illustrated by his/her co-author role, thus focussing on his/her  joint publications, - and their impact through citations? Whence, on one hand, this paper very briefly examines bibliometric laws, like the $h$-index and subsequent debate about co-authorship effects,  but  on the other hand, proposes a measure of collaborative work through a new index. Based on data about the publication output of a specific research group, a new bibliometric law is found.   

 Let a co-author  $C$  have written  $J$  (joint) publications with one or several colleagues. Rank all the co-authors of that individual according to their number of joint publications, giving  a rank  $r$ to each co-author,  starting with $r=1$ for  the most prolific. 
 It  is empirically found that   a very simple relationship holds between the number of joint publications $J$ by coauthors and their rank of importance, i.e. $J \propto  1/r$. Thereafter, in the same spirit as for the Hirsch core, one can define a "co-author core", and introduce  indices operating on an author. It is emphasized that the new index has a quite different (philosophical) perspective that the $h$-index. In the present case, one focusses on  "relevant" persons rather than on "relevant" publications.

Although the numerical discussion is based on one case, there is little doubt that  the law can be verified in many other situations. Therefore, variants and generalizations  could be later produced in order to quantify co-author roles, in a temporary or long lasting stable team(s), and lead to criteria about funding, career measurements or even induce career strategies.
\end{abstract}

  
\maketitle

\begin{table}[t]
\vskip 3.5cm
\begin{center}
\begin{tabular}{|c|c|c|}
\hline &&\\ 
 &   gives the number  distribution/probability&   $n_r \sim n_1/r^{1+\alpha}.$  \\
 Lotka-Pareto  law                                          &of  scientists as a function of & 
 $\alpha\;\sim\; 1$; $n_x = C/(1+ x)^{1+\alpha}$ \\
                           &the  number of  papers they wrote&    $p(x) = \frac{\alpha}{x_0} \left(\frac{x_0}{x} \right)^{\alpha + 1}$ \\
&&\\ \hline
  \hline &&\\ 
  &  asymptotically&  $p(x) = \frac{\mu}{\lambda}B \left(x, \frac{\mu}{\lambda} + 1 \right) = 
\alpha B (x, \alpha +1),$  
 \\
  Yule   distribution  &  corresponds to& where $B(x, \alpha + 1) = \Gamma(x) \Gamma( \alpha x + 1)/ \Gamma(x + \alpha + 1)$, \\
                           &  Lotka  law& $i.e., p(x) \propto \Gamma(\alpha + 1)\alpha/x^{1+ \alpha}$ \\
 &&\\  \hline   
 \hline &&\\ 
   &       ranks scientists&$C = n_1$, one has  $x_r = n_1/(r + a)$
     \\ 
       Zipf-Mandelbrot law  &  by the number of papers&$x_r = \left(\frac{A}{r + B}\right)^{\gamma}$
       \\   
                    &      they wrote& $A = (C/\alpha)^{1/\alpha}; \hskip.25cm B = C/(\alpha 
k_{\max}^{\alpha} ); \hskip.25cm  \gamma = 1/\alpha.$  \\
&& \\  \hline
  \hline  &&  \\
    &      reflects the fact that most of the productivity  $R(n)$ &\\
  Bradford law                    &      of relevant articles by scientists are concentrated&$R(n) = n_1 \ln \left( \frac{n}{a} +1 \right).$\\
                    &      in  a small number $n$  of journals&\\ 
&& \\ \hline
\end{tabular}
\end{center}
\caption{Bibliometric laws  with  a few  words on their origin and/or usefulness; for more details see Sect. 6.1-6.2 in \cite{ch3}}\label{laws}
\end{table}

  \section{Introduction}\label{intro}
    
 In 1926, Lotka \cite{lotka26} discovered a "scientific productivity law", i.e. the number   $n_a$  of authors who has published $p$ papers,   in some scientific field, e.g. chemists and physicists \cite{lotka26}, approximately  behaves like
 
\begin{equation}\label{lotka1}
n_a \sim N /p^2.
\end{equation}  
 where $N $ is the number of authors having published only 1 paper  in the examined data set.
 The $y$-axis can be turned into a probability (or "frequency") by appropriate normalization.
 Conversely, $r-$ranking  authors as 1, 2, 3,...$n_r$  according to their number $n_p$ of publications,  one approximatively  finds: $n_r \sim N_1/r^2$, where   $N_1 $ is the number of  papers written by the most prolific author ($r=1$)  in the examined data set. - a  "ranking law" similar to those discussed by Zipf \cite{Zipf2}.  
However, Lotka's inverse square law, Eq.(1),  is only a theoretical estimate of productivity:  the square law dependence is not always obeyed \cite{lotkalaw?MLP}. In fact,   author inflation leads to a breakdown of Lotka's law   \cite{KRETSCHMERROUSSEAU Lotka,Egghebook05,EggheRousseauLotka} 
 the more so nowadays. 
 It is known, see $http://www.improbable.com/airchives/paperair/volume15/v15i1/v15i1.html$,  that  a collaboration of large team effect has led to articles with more than 2000 authors, e.g. in  physics or in medicine. 
 The roles or effects of coauthors is thus to be further studied in bibliometrics.

  Note that several other so called  $laws$ have been predicted or discovered  about relations between number of authors, number of  publications, number of citations,  fundings, dissertation  production, citations, or the number of  journals or scientific  books,  time intervals, .... etc.   \cite{yab1,ch3,price63,price75,price78,gilbert,chung,kealey}.  Scientometrics has become a scientific field in itself \cite{GLANZELCOURSEHANDOUTS2003,beck,Potter88,vanRaan,egg1,andrea1}.
 Thus,  statistical approaches and bibliometric models  based on the laws  and distributions of Lotka, Pareto,  Zipf-Mandelbrot, Bradford,  Yule, and others, - see Table \ref{laws} for a summary, do provide much useful information for the analysis  of the evolution of scientific systems  in which a development is closely connected to  a  process of idea diffusion and work collaboration. Note that the "laws" do not seem to distinguish single author papers from collaborative ones. Lotka, himself, assigned each publication to only one, the senior author, ignoring all coauthors. Thus, again,  coauthorship raises questions. .
 
 More recently,   an index, the $h-$index, has been proposed in order to quantify an individual's scientific research output \cite{hindex}.  A scientist has some index $h$, if $h$ of his/her papers have at least $h$ citations each. {\it A priori} this $h-$ value is based on journal articles. Note that, books, monographs, translations, edited proceedings, ...  could be included in the measure, without much effort.    Such a "measure"   should obviously be robust and should not  depend on the precision of the examined data basis. No need to say that  the official publication list of an author should be the most appropriate starting point, - though the list should be reexamined for its veracity, completeness  and  appropriateness. 
However, it is rather unusual that an author  records by himself the citations of his/her papers. In fact, sometimes, several citations go unnoticed.  Alas, the number of citations is also known to vary   from one search engine to another, - even within a given search engine, depending on the inserted keywords \cite{Buchanan}.  Thus much care must be taken when examining any publication and its subsequent  citation list. This being well done,  the   {\it  core of a publication list} is a notion which can be defined, e.g., as being the set of papers which have more than  $h$ citations.  Fortunately, this "core measure" does not vary much,  whatever the investigated data base, - because of the mathematical nature of  the functional law, a  power law, from which the index  is derived \cite{hindex,Egghebook05}.

Let it be noted that a discussion focusing on the many  defects and improved variants of the $h$-index, e.g. the 
$b-$, $e-$, $f-$, $g-$,  $hg-$, 
$m-$, $s-$, $A-$, $R-$, ... indices, their computation and standardization can be found in many places, e.g. in 
 \cite{AlonsoetJOI3.09,ARIST44.10.65LEgghe,Radiology255.10.342 bibliom,Schreiber2010b,Schreiber2012JoI6_17v}.   
 It can be emphasized that these indices  are more $quantity$ than $quality$ indices,    because they operate at the level  of the paper  citation number, considered to be relevant for measuring some author visibility     \cite{0912.3573 zhang}.
     Also, one can distinguish between   "direct indices" and "indirect indices"  \cite{JASIST59.08.830BornmannMutzDaniel}. 
     It can be also  debated that inconsistencies might arise  because of self-citations \cite{Schreiber2007EPL}, though such a point is outside the present discussion.  However,  any cited  paper is  usually considered as if it was written  by a single author. Nevertheless, 
  there can be multi-authored papers.    
It is clear, without going into a long discussion, that the role and the impact  of such co-authors are difficult to measure or even estimate 
  \cite{Laudel,BEAVER52.01}.  
 One may even ask  whether there are sometimes too many co-authors  \cite{McDK95}.  
 
In order to pursue this discussion, a brief review of  the literature on collaborative effects upon the $h$-index is presented in Sect.  \ref{hreview}. This will  serve as much as   to present a framework for  the state of the art on such $h$-index spirit  research, taking into account collaborative aspects, together with quick comments, as  suggesting arguments  on  the interest of the present report, - henceforth  justifying this new approach.
  
 The present paper is an attempt to $objectively$ quantifying the importance of co-authors, whence {\it a priori} co-workers  \cite{0354-73100303211V}, in scientific publications, over a "long" time interval,  and consequently  suggesting further investigations about their effect  in  (and on) a team.  In other words, the investigation, rather than improving or correcting the $h$-index and the likes, aims at  finding a new structural index which might "quantify" the role of an author as the leader of co-authors or coworkers.  It will  readily appear that  the approach can automatically lead to criteria about,  e.g., fundings, team consistency, career "measurements", or even induce career strategies. 

As a warning, let it be mentioned that the present investigation has not taken into account the notion of  network. Of course, every scientist, except in one known case, has published with somebody else, himself/herself having published with  some one  else, etc.;  a scientific network exists  from the coauthorship point of view. But the structure and features of such a huge network can be found in many other publications   \cite{HK94,HK97,Sigmod03,1006.0928 criticalmass kenna,HKretschmer85structuresize,HKUKTK07}, to mention very few.   

Since the size and structure of a temporary or long lasting group  is surely relevant to the productivity of an author \cite{1-s2.0-S1751157711000630-main}, this will be a parameter   to be still investigated when   focussing on network nodes or fields, as  in the  investigated data here below. However, the goal is not here to investigate the network, but rather concentrate on the structural aspect  of  the neighborhood of one selected node. It should not be frightening that a  finite size   section of the network only is investigated. It is expected that the features found below are so elegant, and also simple, that they are likely to be valid  whatever the node selection in the  world scientific network \cite{MelinPersson96,PNAS101.04-Newman-5200-5 coll netwk,Zuccala06invisblcoll,ch6}.
  
 Firstly, an apparently not reported "law", quantifying some degree of research collaboration \cite{1-s2.0-S1751157711000800-main}, is presented, in Sect. \ref{data}.  A very simple relationship is empirically found, i.e., the number of joint publications $J$ of a researcher with his/her co-authors $C$  and their rank $r_J$  (based on their "publication frequency") are related by $J \propto 1/r_J$, - a simple law as the second Lotka law, linking   the number of citations $c_p$ for a publication $p$ of an author and their rank, $r_p$ according to their number of citations,   $c_p\propto 1/r_p$, i. e., leading to the $h-$ index \cite{hindex}. 

Secondly, in Sect. \ref{core}, one  defines {\it  the core of co-authors for an individual}, - it is emphasized that this measure has a quite different spirit that the definition of {\it  the core of papers for the publication list of an individual}, in the $h-$ index scheme.   
 Numerical illustrations are provided. Some analysis of the findings and some discussion   are found in Sect. \ref{analysisdiscussion}. A short conclusion is found in Sect. \ref{conclusion}.

\section{ $h$-index: Collaborative effects, - a few comments to serve as  a brief review}\label{hreview}

As mentioned here above, inconsistencies have been shown to  arise  on the $h$-index when one takes into account multi-authored papers  \cite{Vanclay}.     Several  disturbing, or controversial, effects of multi-authorship on citation impact, for example, have been shown in bibliometric studies by Persson et al. in 2004   \cite{Perssonetal04}. Yet, Gl\" anzel and  Thijs \cite{WGBT04} have shown that  multi-authorship does not result in any exaggerate extent of $self-citations$.  In fact, self-citations can indicate some author creativity,  or versatility at changing his/her field of research \cite{Ausloos08,Iina06,Iina07,Iinamadrid}

To take into account the effect of multiple $co-authorship$ through the $h-$index,  Hirsch 
 \cite{hirsch10},  himself, even proposed the $\hbar$  index     as being  the number of papers of an individual that have a citation count {\it larger     than or equal to} the $h-$index of  {\it all  co-authors} of each paper.   Of course,  $\hbar \leq h$. With the original $h$-index a multiple-author paper in general belongs to the $h$-core of some of its coauthors and not belong to the $h$-core of the remaining coauthors.  The $\hbar$- index, unlike the $h$-index, uniquely characterizes a paper as belonging or not belonging to the $\hbar$-core of its authors. However, these considerations emphasize "papers" rather than "authors". Indeed, one focusses on some "paper-core", not on some  "co-author-core".
    
   It has been much discussed whether co-authors must have all the same "value" in quantifying the "impact" of a paper; see \cite{ioannidis,0354-73100303211V}, and also \cite{AJR167.96.slonecoauthor} pointing to "undeserved coauthorship".

For a practical point of view, Sekercioglu proposed that the $k-$th ranked co-author be considered to contribute $1/k$ as much as the first author \cite{Sekercioglu08},  
 highlighting an earlier proposal by Hagen \cite{Hagen09}.  At the same time, Schreiber proposed  the $hm-$index \cite{Schreiber2008d,Schreiber2008c},   and $g(m)$ index \cite{JoI4.10.42Schreibergm}, counting the papers equally fractionally according to the number of authors; see also  Egghe \cite{Egghe2008d} giving an author of an $m-$authored paper only a credit of $c_p/m$, if the $p$ paper received $c_p$ citations.  Carbone \cite{Carbone} recently also proposed to give a weight  $m^{\mu}_i$	to each $i$-th paper of the $j-$th individual according to  the number $m_i$ of co-authors of this $i$-th paper, - $\mu$ being a parameter at first. Carbone argued that   ambiguities in the  e.g. $h-$ index  distribution of scientist populations are  resolved if  $\mu\simeq1/2$. Other considerations can be summarized : (i)    Zhang   \cite{ZhangEMBO10.09}  has argued against Sekercioglu hyperbolic weight distribution,  as missing the corresponding author, often the research leader. Zhang  proposed that weighted citation numbers, calculated by multiplying regular citations by weight coefficients, remain the same as regular citations for the first and corresponding authors, who can be identical, but decreased $linearly$ for authors with increasing rank; (ii) Galam \cite{galamscientom12} has recently proposed another fractional allocation scheme for contributions to a paper,  imposing in contrast to Zhang, that    the total weight of a paper equals 1, {\it in fine} leading to a $gh-$index favorizing a "more equal" distribution of "co-author's weight" for more frequently quoted papers.     Note that it differs from the  $hg-$index; see  \cite{AlonsoetJOI3.09}.  

Other considerations have been given to  the co-authorship "problems".   Nascimento et 	al.  \cite{Sigmod03}   e.g. found out  that co-authorship is a small world network.   from such a point of view,  B\" orner et al. \cite{KBLDWMKAV05}, but also others    \cite{ivs.2009.31}, used a  {\it weighted graph} representation  to illustrate the number of publications and their citations. However,   even since  
Newman \cite{PNAS101.04-Newman-5200-5 coll netwk} or more recently  Mali et al. \cite{ch6}  and the subsequent works here above recalled,  there have been considerations on the number   of co-authors and their "rank", for $one$ paper among many others of an individual, but no global consideration in the sense of Hirsch, about "$ranking$" all coauthors, over a whole  joint publication process.

 Author productivity and geodesic distance in co-authorship networks, and visibility on the Web, have also been considered to illustrate the globalization of science research \cite{HKretschmer04collaborationgeodesic,1-s2.0-S1751157711000599-main,Ponds}.

\begin{table} 
\begin{center} 
\begin{tabular}[t]{lclccc|cc} 
\hline 
 
& \phantom{fBm}MA & \phantom{fBm}PC  & \phantom{fBm}AP& \phantom{fBm}JP & \phantom{fBm}JK & \phantom{fBm}TK& \phantom{fBm}DS \\ \hline\hline 
 &&&&&&& \\  
born in&\phantom{fBm}1943&\phantom{fBm}1945&\phantom{fBm} 1937& \phantom{fBm} 1939&\phantom{fBm}1939   &\phantom{fBm}1972&\phantom{fBm}1943
\\Ph.D. in&\phantom{fBm}1973&\phantom{fBm}1973&\phantom{fBm} 37& \phantom{fBm} none&\phantom{fBm}1973 & \phantom{fBm} 2001.&\phantom{fBm}1970\
\\tenure in&\phantom{fBm}1986&\phantom{fBm}1976&\phantom{fBm} 1980& \phantom{fBm} none& \phantom{fBm}1995 &\phantom{fBm}  2007.&\phantom{fBm}1977
\\1st publication  in&\phantom{fBm}1971&\phantom{fBm}1974&\phantom{fBm} 1966& \phantom{fBm} 1983& \phantom{fBm}1967 &\phantom{fBm}1997&\phantom{fBm}1967
\\latest recorded publication  in&\phantom{fBm}2010&\phantom{fBm}2010&\phantom{fBm} 2010&\phantom{fBm} 1983&\phantom{fBm}1999 & \phantom{fBm}2010& \phantom{fBm}2010 
\\  &&&&&&& \\   \hline  
 &&&&&&&\\ $n_p$: Numb. publications ($<$2011)  &\phantom{fBm} 571&  \phantom{fBm} 34&\phantom{fBm} 111&\phantom{fBm}2&\phantom{fBm}60&\phantom{fBm}38&\phantom{fBm}638\\ 
$J$: Numb. joint publications ($<$2011) &\phantom{fBm} 528&  \phantom{fBm} 33&\phantom{fBm} 90&\phantom{fBm}2&\phantom{fBm}48&\phantom{fBm}38&\phantom{fBm}486\\ 
$s$: Numb. single author publications ($<$2011) &\phantom{fBm} 43&  \phantom{fBm} 1&\phantom{fBm} 21&\phantom{fBm}0&\phantom{fBm}12&\phantom{fBm}0&\phantom{fBm}152\\ 
Numb. jointly ed. books ($<$2011)&\phantom{fBm} 9&\phantom{fBm} -  &\phantom{fBm} 8&\phantom{fBm}--&\phantom{fBm}(2; transl.)&\phantom{fBm}--&\phantom{fBm}10$^*$\\   &&&&&&& \\ \hline  &&&&&&&\\

 \phantom{fBm} $h-$index \cite{hindex} &\phantom{fBm}35&\phantom{fBm} 11&\phantom{fBm} 10&\phantom{fBm} 2&\phantom{fBm} 10&\phantom{fBm}6&\phantom{fBm}55\\
 Numb. cit. of most often cited paper &\phantom{fBm}152&\phantom{fBm}127&\phantom{fBm} 37& \phantom{fBm} 7& \phantom{fBm}537&\phantom{fBm}41&\phantom{fBm}1430\\  
 Tot. numb. citations till $h$&\phantom{fBm}1113&  \phantom{fBm}296 &\phantom{fBm}224&\phantom{fBm}14&\phantom{fBm}745&\phantom{fBm}100&\phantom{fBm}8148\\
\phantom{fBm} $A-$index \cite{Jin06} &\phantom{fBm}31.8&\phantom{fBm} 26.9&\phantom{fBm} 22.4&\phantom{fBm} 7&\phantom{fBm} 74.5 &\phantom{fBm}16.7&\phantom{fBm}148.1\\ &&&&&&& \\
  \hline
  \hline 
  &\phantom{fBm} &\phantom{fBm} &\phantom{fBm} & \phantom{fBm}  &\phantom{fBm} &\phantom{fBm}  &\phantom{fBm} 
\\Numb. of diff. co-authors  ($r_M)$&\phantom{fBm}317&  \phantom{fBm} 32&\phantom{fBm} 46& \phantom{fBm} 4& \phantom{fBm} 38&\phantom{fBm}51&\phantom{fBm}285\\ 
Total numb. of  co-authors $\equiv$  $\Sigma_{i=1}^{n_p} \Sigma_{j=1}^{r_M}  \;J_{ij}$ &\phantom{fBm}1551&  \phantom{fBm} 95&\phantom{fBm}134& \phantom{fBm} 8& \phantom{fBm} 108&\phantom{fBm}181&\phantom{fBm}793\\ 
 Tot. coauthor distribution skewness&\phantom{fBm}7.35&  \phantom{fBm}4.66&\phantom{fBm}3.18& \phantom{fBm} -& \phantom{fBm} 2.18&\phantom{fBm}3.39&\phantom{fBm}3.98\\  &&&&&&& \\\hline
  &&&&&&&\\
\phantom{fBm} $m_a-$index &\phantom{fBm}19&\phantom{fBm}   4&\phantom{fBm} 7&\phantom{fBm} 2 &\phantom{fBm} 5&\phantom{fBm}6&\phantom{fBm}12\\
Numb. J. Publ. with "$best$" co-author &\phantom{fBm}155&  \phantom{fBm} 30&\phantom{fBm} 21& \phantom{fBm} 2& \phantom{fBm} 13&\phantom{fBm}26&\phantom{fBm}30\\ 
Numb. J. Publ till $m_a$ $\equiv$ $\sum_{i=1}^{m_a}\; J_i $ &\phantom{fBm}810&\phantom{fBm}46&\phantom{fBm} 170&\phantom{fBm} 4&\phantom{fBm} 39&\phantom{fBm}76&\phantom{fBm}264\\
\phantom{fBm} $a_a-$index, Eq.(\ref{a_a}) &\phantom{fBm}42.6&\phantom{fBm}11.5&\phantom{fBm} 24.3&\phantom{fBm} 2&\phantom{fBm} 7.8&\phantom{fBm}12.7&\phantom{fBm}22\\ &&&&&&& \\\hline   &&&&&&&\\
\phantom{fBm}$m_a /r_M$&\phantom{fBm}0.06&\phantom{fBm}0.125&\phantom{fBm} 0.15&\phantom{fBm} 0.50&\phantom{fBm} 0.13&\phantom{fBm}0.12&\phantom{fBm}0.04  \\ 
\phantom{fBm}$a_a /r_M$&\phantom{fBm}0.13&\phantom{fBm}0.36&\phantom{fBm} 0.53&\phantom{fBm} 0.50&\phantom{fBm} 0.21&\phantom{fBm}0.25&\phantom{fBm}0.08  \\ &&&&&&& \\\hline   \hline

\end{tabular} 
   \caption{ Data reduced from CV or Google Scholar on hereby examined scientist set. Books by JK are translated from english (into polish); $^*$: DS is editor of a book series not counted in the 10}\label{dataMAPCAPJPJK}
\end{center} \end{table}

\section{Data  }\label{data}

\subsection{Methodology }\label{Methodology}

The main points of the paper are based on a study of the publication list of a research group, i.e. the SUPRATECS Center of Excellence at the University of Liege, Liege, Belgium, at the end of the 20-th century.  The group was involved in materials research, and involved engineers, physicists and chemists.  Among the researchers, a group of  5 authors (MA,PC,AP,JP,JK) has been selected for having various scientific careers,  similar age, reasonable expertise or reputation,  with an expected sufficient set of publications and subsequent citations, spanning several decades, thus allowing   to make some  acceptable statistical analysis. These 2 females and 3 males are part of a college subgroup of the SUPRATECS, having mainly performed research in theoretical statistical physics, but having maintained contacts outside the Center, and performed research on  different topics; e. g., see some previous study on AP can be found in   \cite{Ausloos08}.   

Each publication list,  as first  requested and next kindly made available by each individual, has been manually examined, i.e. crosschecked according to various search engines and different keywords,   for detecting flaws, "errors", omissions or duplications. Each investigated list has been reduced to joint papers published in refereed journals or in refereed conference proceedings.  A few cases of "editorials" of conference proceedings have been included, for measuring the $h$-index, see next subsection. Sometimes such citations exist, instead of the reference to the book or proceedings. But these do not  much impair the  relevant numerical analysis of the new index.

Finally, in order to have some appreciation of the robustness of the subsequent finding, a test has been made with respect to two other  meaningfully different scientists, so called "asymptotic outsiders'' :  (i) the first one,  TK, is a younger female researcher , an experimentalist, sometimes having collaborated with the group, but outside statistical mechanics research, - a chemist,  known to have several, but not many,  joint publications with the main 5 individuals; (ii) the  other, DS, a male, is a   well known researcher, of the same generation as the 5 main investigated ones; DS is known as a guru in the field, has many publications, many citations, thus has expectedly a larger $h$ -index, and is known to have published under  "undeserved co-authorship".  He is also chosen, as a test background, because having very few joint publications  with the 5 main investigated authors, but has a reliable list of published works.

A brief CV and the whole list of publications  of  such 7 scientists are available from the author.  Note that it is somewhat amazing that for such a small number of authors, a hyperbolic Lotka-like law is verified with a $R^2\simeq 0.995$, though the exponent is close to 2.9 (graph not shown). 

\subsection{$h$-index  and relevant  bibliometric data of the investigated scientists} \label{5h}
 
In order to remain in the present bibliometric framework, the $h$-index of such (7) scientists  has been manually measured.  
Care has been taken about the correctness of the references and citations. For example, JP and DS have a homonym in two other fields.  Also, the total citation count till the end of the examined time interval, i.e., 2010, has not been possible for MA, AP and DS, due to their rather long publication list.  The citation count has been made up to their respective $h$-index, - to measure their $A$-index. But, it is  emphasized that the citation count is irrelevant for measuring the presented new index below. In Table \ref{dataMAPCAPJPJK},  the number of citations, leading to the $h$-index value, includes books when they are recorded as papers in the search engines, papers deposited on arXives, and papers published in proceedings,  be they in a journal special issue or in a specific book-like form. 

Note that for the $h$-index, Google Scholar distinguishes the citations for a paper uploaded on the arXives web site and a truly published paper. For the present investigation on joint publications, both "papers",  obviously having the same authors, are considered as only one joint publication!

The  $h-$index, with other relevant data, like the  total number of  publications, or the number of joint publications for which they are co-authors are given in Table \ref{dataMAPCAPJPJK}.  Let it be emphasized that the joint publications have covered different time spans. These publications, of course, involved other co-authors than those  selected 5 members.

Also recall that the number of "citations till $h$" when divided by $h$ is equal to the $A-$index, measuring   the  reduced area of the histogram  till $r=h$, i.e. $A =(1/h)\sum_{p=1}^h c_p$. \cite{Jin06}.

  \begin{figure}
\centering
\includegraphics[height=18cm,width=15cm]{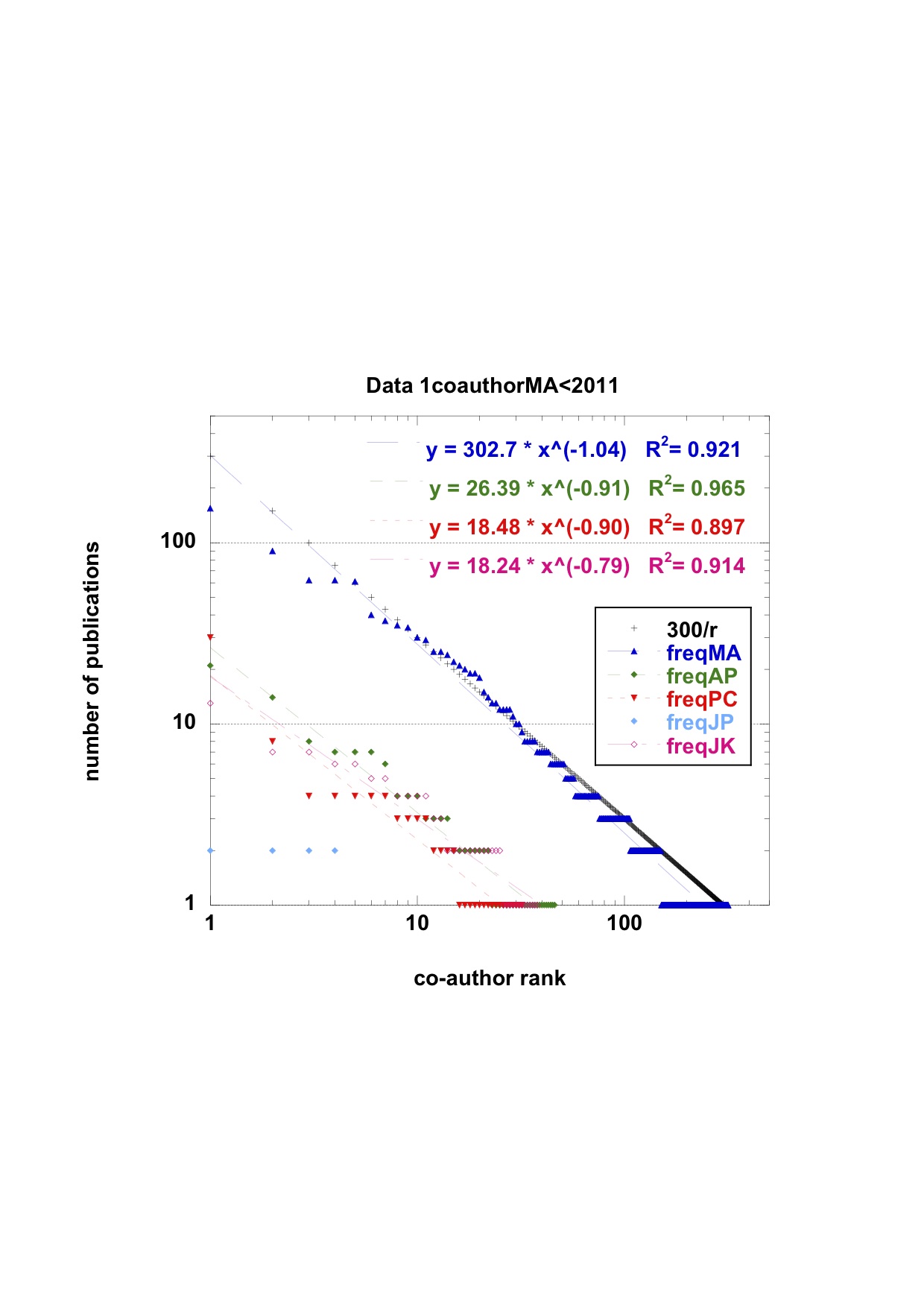}
 \caption{ 
log-log plot of the number of (joint) publications with coauthors ranked according to rank importance, for  the 5 team members; a few power law lines are indicated; the $J\;\simeq 300/r$ law is given as a guide to the eye}
\label{Plot44c1MAAPPCJPJK4fit}
\end{figure}

\begin{figure}
\centering
\includegraphics[height=18cm,width=15cm]{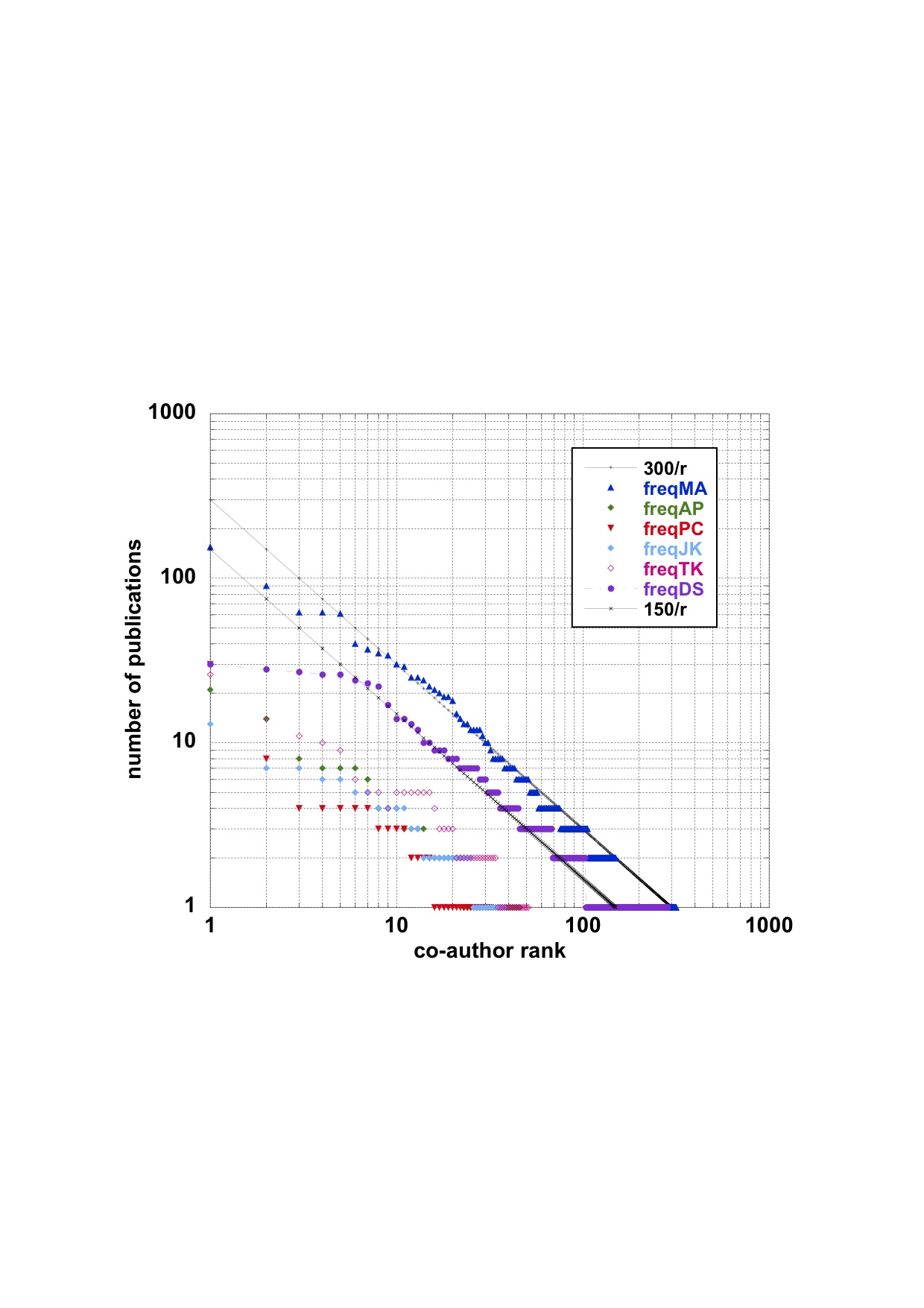}
\caption{ 
log-log plot of the  number of (joint) publications with co-authors  ranked according to importance for  the examined  team members and outsiders ;  (-1) power law lines are indicated for the two most prolific authors, MA and DS. Note the well marked curvature at low rank}
\label{Plot111all-1}
\end{figure}

\subsection{Zipf plots of Joint Publications vs. Co-author Ranking} \label{Zipfplots}

Having established the number of joint publications $J$ of the  (5+2) scientists here mentioned,  and ranked them according to the frequency of joint publications with one of the main authors, the most usual graph to be done  is a Zipf plot \cite{Zipf2,httpZipf,li02}.  
By an abuse of language, the number of joint publications is  called $freq$ for frequency. However, no scaling has been made with respect to the total number of publications of each author.   First,
a log-log plot of the number   of joint publications  between the five team members,  with whoever partners ranked in a decreasing order of joint contributions, is given in Fig. \ref{Plot44c1MAAPPCJPJK4fit}. The data appears to fall on a straight line, with a slope equivalent to a power law exponent   $\simeq -1$, i.e. 
\begin{equation}\label{ZlikeCr}
  J\propto \frac{1}{ r }.
  \end{equation}
    Note that the data appears better to fall more  on a line (on such a plot),  if the number of publications of the authors is large.   Such a Zipf behavior does not seem to have been reported in this bibliometric context \cite{httpZipf,li02}. 
    
    In view of such data, one of the  investigated scientists, JP, has been removed below from  the plots for better clarity;  in fact, JP has  peculiar characteristics, since this researcher has no  Ph.D.  and has not continued  publishing, after participating in the  SUPRATECS activities.

A  comparison with the "two asymptotic outsiders", TK  and DS, can next be made,   as a test of the scientific field, sex and age (ir-)relevance, when obtaining the above "law". A log-log plot of the number of joint publications versus ranked co-authors, be they partners or not, ranked in a decreasing order of joint contributions, is   given for these 6 authors, in Fig. \ref{Plot111all-1}.  The power law exponent is emphasized to be $very$ close to $-1$ particularly for the most prolific authors. Nevertheless, one may observe a curvature at "low rank". This indicates that a Zipf-Mandelbrot-like form

\begin{equation}\label{ZMlikeCr}
  J =\frac{J_1}{(n+r)^\zeta},
  \end{equation}
  with $\zeta\simeq1$,
  would be more appropriate. 
  This is a very general feature of almost all Zipf plots \cite{httpZipf,li02}. 
  
 A few interesting features have to be observed,  at low rank. The first points, here even the first two points for MA, sometimes surge up from the Zipf-Mandelbrot-like form. This is  known as a "king effect" , i.e.  the rank = 1 quantity is much larger  and is much above the straight line on such plots. This is  for example the case of country capitals when ranking cities in a country; e.g. see Fig. 7 in \cite{EPJB2.98.525stretchedexp_citysizesFR}. Here, the surge up indicates the importance of the main pair of authors, relative to the others.  On the other hand, a (obviously to be called)  "queen effect" occurs when the low rank data falls almost on a horizontal, - as for DS. The interpretation is as easy as for the king effect: several authors,  always the same ones, have some disposition toward the "queen", in  terms of joint publications. Some observation of this feature related to careers will be discussed below.
  
The  best fits  by a power law  and by a  Zipf-Mandlebrot law, Eq.(\ref{ZMlikeCr}) of the number of publications versus co-author rank are given in Table  \ref{fitdata},   for the 4 main  researchers  and the 2 outsiders.  
  The fit with the latter law is of course much more precise, though one might argue that this is due only to the number of involved parameters. The fit parameters, given in  Table \ref{fitdata}, nevertheless indicate some universality in behavior.

 \begin{table}
           \begin{center}
           \begin{tabular}{|c|c|c|c|c|c|c| c| c| c| c|  }
\hline   $authors$  &&$n $&$J_1$&$\zeta$ &$R$ &&$J_1$&$n$ &$\zeta$ &$R$   \\ 
\hline MA&&0&302.7&$ 1.042$&$ 0.960$  &&413.78 &1.5&$1.101$&0.9927 \\    
\hline AP&&0&  26.39&$ 0.911$&$0.983$  &&37.809 &0.8&$1.011$& 0.9911  \\   
\hline PC&&0&18.48&$0.904$&$0.947$  &&8.7187 &-0.9&$0.6526$ &0.9926 \\ 
\hline JK&&0&18.24&$0.787$&$0.956$  &&49.172 &2.55&$1.068$  &0.9746  \\   
\hline TK&&0&34.77&$0.897$&$0.982$  &&48.495&0.7&$0.9879$& 0.9860\\   
\hline DS&&0&76.96&$0.822$&0.864&&159.88 &$5.6$  &0.9599& 0.9838  \\   
\hline
 \end{tabular} \label{fitdata}
    \end{center} \caption{Pertinent values of fit parameters, and fit precision ($R$), to the hyperbolic form mimicking the theoretical second Lotka law,   i.e. when  $n=0$  in the  Zipf-Mandlebrot law, Eq.(\ref{ZMlikeCr})  } \end{table}

 \section{ Co-authorship core:  the $a-$ and $m_a-$ indices}\label{core}
 
\begin{figure}
\centering
\includegraphics[height=18cm,width=15cm]{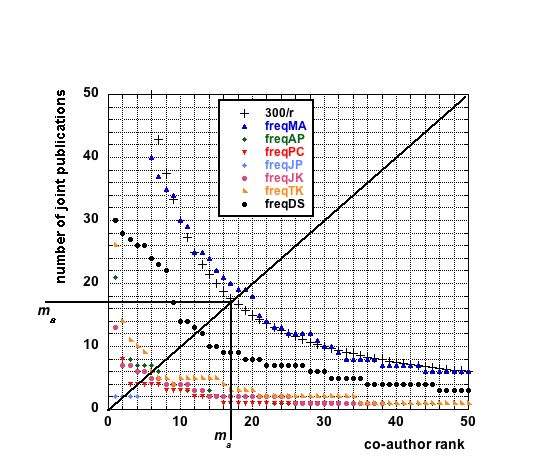}
 \caption{  Plot of the number of (joint) publications with coauthors ranked according to the number of contributions of  a given coauthor; the rank decreases with the coauthor importance;  this allows to define the core of co-authors for a given author through an index $m_a$; values are given in Table II}
\label{fig:core}
\end{figure}

 The Hirsch core of a publication list is  the set of publications of an author which is cited more than $h$-times.
 Similarly to the definition of the  "Hirsch core",  along the $h-$index, or also the $\hbar$-index, concept, one can define the {\it core of coauthors for an  author}. This value, here called $m_a$,  is easily obtained from Fig. \ref{fig:core}, in the cases so examined through a simple geometrical construction.  It would then be easy to obtain, e.g. from the publication list or the CV, who are the main partners of the main coauthors, and  do make more precise the active members of a team.

Similarly to the $A-$ index \cite{Jin06}, one can define an   $a_a$- index which measures the  surface below the empirical data of the number of joint publications till the coauthor of rank $m_a$, i.e.
\begin{equation}\label{a_a}
  a_a =\frac{ 1}{m_a} \sum_{i=1}^{m_a} J_{i}.
  \end{equation}
  In practical terms,  it is an attempt to improve the sensitivity of the $m_a-$index to take into account the number of  co-authors  whatever the number of joint publications among  the most frequent coauthors.
 The results are given in Table II.

 \section{Some analysis and some discussion} \label{analysisdiscussion}
 
  \subsection{ Analysis  } \label{analysis}
 Three subtopics are to be commented upon. First, the high quality of the fits  can be noticed. More precisely,  the Zipf-Mandelbrot fit  parameters allow to distinguish authorship patterns. For example, the parameter values  indicate that PC and DS are markedly different authors as their coauthors are concerned. 
  It should be remarked that the  most anomalous  parameters occur  for PC, for which $n$, in Eq.(\ref{ZMlikeCr}),  is negative ($n\simeq -0.9$), together with a low $\zeta$ ($\simeq 0.65$). From Table II, it is observed that PC has not only  the lowest number of  publications as well as the lowest number of   joint publications,  but also the relative highest number of coauthors. In contrast DS has a large $n$ parameter ($\simeq 5.6$), with $\zeta\simeq 1$. 

Next, the $h-$index values  show that such authors can be grouped in three sets, MA \& DS, PC, AP \& JK, TK \& JP. Observe that JK has nevertheless a very high number of citations for one paper, leading to a large $A$-index ($\simeq 74.5$), about half of that of DS  ($\simeq 148$). Except for JP, all other authors  can be grouped together from the $A-$index point of view, and approximately conserve the same ranking as for the $h-$index.

Finally, concerning the new measure of the scientific productivity through joint publication coauthorship, one may consider that, at least one can group coauthors  in two regimes: those with a $r$ small, thus frequent, likely long standing truly coworkers or group leaders ($\sim$ $"bosses"$), and those with a $r$ large, most often not frequent coworkers, who are likely  students, post-docs, visitors, or sabbatical hosts types. The threshold, 	according to Fig. 2, occurs near $r\sim 10$, which seems a reasonable  value of the number of "scientific friends". However, the $m_a$ index gives a more precise evaluation of the core of coauthors for a given scientist.  Table III indicates that there is marked difference between the MA \& DS type of scientists, and the others from the point of view of association with others. Their $m_a$-index is quite above 10, indeed. It seems that one might argue also that the importance of leadership (or centrality using the vocabulary of network science)   might be better reflected through the area of the histogram of joint publications with the "main" coauthors through the $a_a$-index. In so doing, one obtains a high value ($\simeq 43$) for MA, while AP and DS fall into a second group with $20 \leq a_a \leq 25$.   The analysis therefore leads to  suggest that one has thereby obtained a criterion for indicating either a  lack of leadership, exemplified by PC and TK, or a "more central" role, e.g. for MA, AP, and DS. 
   
 \subsection{Discussion} \label{discussion}
 
  The interpretation of such  results indicates the relative importance of working in a team, or not, as well as the sociability or capacity for attraction of some author; see   \cite{LeeBozeman}. Also recalling that the data is a snapshot  at some time of a list of publications,  it points toward further studies on time effects.  Not only the evolution of the list should be relevant in monitoring scientific activities, but also the origin of the time interval  seems to be  a relevant parameter.  Indeed, compare the $m_a$ or  $a_a$ values for  PC and TK, and observe their relative scientific career output as co-authors.  They have a similar record of publications; they started their career at  very different times.    However, TK , though being associated in a temporary but regular way with the SUPRATECS team, though permanently based in another group, has almost the same $a_a$ (slightly greater than 10)  as PC, a stable senior partner for SUPRATECS. Even though PC has many less co-authors.  Be aware that TK is an experimentalist and PC a theoretician, - both females. 
  
  Similarly, compare $a_a$ for AP and DS, both with $a_a$ above 20, even though their number of co-authors is markedly different, - 32 vs. 285.  Recall that   in 40 years, the average number of authors of scientific papers has doubled. Some say the problem reflects the growing complexity of the research process in many disciplines 
  \cite{McDK95}. 
However, the starting career time is about the same. 
To compare, -if necessary, the role of coauthors, and some sort of leadership in a scientific career, therefore, an indirect measure of interest can be the ratios $m_a/r_M$ or/and  $a_a/r_M$. In so doing, according to Table III, DS and MA are proved to be "leading".   It is rather evident that the inverse of such ratios  are of the order of magnitude of the degree of the author node of the scientific network.  In some sense,  $m_a/r_M$ or/and  $a_a/r_M$ weight the links attached to a node; see \cite{HKretschmer99collaboration2birds}, about flocking effects.

  \subsection{Career patterns } \label{career}

  A brief comment can be made on the king and queen effects seen on Fig.2, in relevance to the type of career of these authors. For example,  the low ranked coauthors of DS  have an equal amount of joint publications. Same for MA, except for the lowest rank coauthor who has a large relevance. These two features point to career hints. One way to interpret this feature can be indeed deduced from Table  \ref{dataMAPCAPJPJK}.  It can be observed that the tenure year markedly differs for both authors, - leaders. It can be understood that  DS had more quickly possibilities of collaborations with selected co-authors than MA who had on one hand to list co-authors of hierarchical importance on joint publications during a longer time, -thus  a "queen effect", and on the other hand had  to rely on an experimentalist ("the king") leader for producing publications.  The same effects are seen for PC and TK on Fig.2. The similarity emphasizes the argument on the role of sex, age and type of activity; see \cite{sexproduct}.

 \section{Conclusion}\label{conclusion}

  Two main findings must be outlined as a summary and conclusion.
 
 \begin{itemize} \item
 A finite set of researchers, from  a large research group, having stable activity, and different types of researchers, all well known to the writer, as been examined. These have been performing and producing papers in theoretical statistical physics.  It has been found that the number of joint publications when ranked according to the frequency a coauthor appears leads to a new bibliometric law: \begin{equation}\label{ma1}
J \propto 1/r.
\end{equation}
The   tail of the distribution seems undoubtedly equal to -1. A deviation occurs for individuals having few co-authors or a limited number of publications. Instead of a $-1$ slope on a log-log plot, one can observe a   $Zipf-Mandelbrot$ behaviour at small $r$, - related to the so called "king effect" and "queen effect". Note that this wording does not  apply to the examined author but to the main co-authors, one "king" at rank =1, the "queens" at rank  $\leq 4$ or so. This leads to imagine a new measure of co-authorship effects, quite different from variants of the$h-$index. The emphasis is not on the number of citations of  papers of an author, but is about how much coworkers he/she has been able to connect to in order to produce (joint) scientific publications.
\item
Next, in the same spirit as for the Hirsch core, one can define a "co-author core", and introduce  indices, like $m_a$ and $a_a$,  operating on an author.
 Numerical results adapted to the finite set hereby considered can be meaningfully interpreted. 
 Therefore, variants and generalizations  could be later produced in order to quantify co-author roles
  in a temporary team. The finite size  of the sample is  apparently irrelevant as an argument against the findings. Nevertheless, one could develop the above considerations, through a kind of network study.
 Of course the present  findings and the proposed indices are only a few of the possible quantitative ways to tackle the co-authorship problem. Different other methods can be investigated, with variants as those recalled in  Sect. \ref {hreview}. However, they will never be the whole answer   to evaluate the career of an individual nor to fund his/her research and team.  But they are easy "arguments" and/or smoke screens.  
 
 \end{itemize}
Thus, it might have been thought that the number of co-authors of papers over a career might be related to the number of joint publications.  But it was not obvious that a simple relationship should be found. In so finding, an interesting new measure of research team leaderships follows, - the  "co-author core". It is hoped that the present report thus can help in classifying scientific types of collaborations \cite{HK87,sonnenwald07}.

 As a final point, let it be emphasized that even though co-authorship can be abusive \cite{Kwok05}, it  should not be stupidly scorned upon. Indeed in some cases,  co-authorship and output are positively related. For instance, it has been shown that, for economists, more co-authorship is associated with higher quality, greater length, and greater frequency of publications  \cite{sauer88,Hollis01}.   Yet bibliometric indicators, as    those  nowadays discussed, can be useful parameters to evaluate the output of scientific research and to give  some information on how  scientists actually work and collaborate.  To measure the quality of the work has still to be discussed.

\begin{flushleft}
{\large \bf Acknowledgment}
\end{flushleft}
  The author gratefully acknowledges stimulating and challenging discussions with many wonderful colleagues at  several meetings of the  COST Action MP-0801, 'Physics of Competition and Conflict'. In particular,  thanks to O. Yordanov for organising the May 2012 meeting  "Evaluating Science: Modern Scientometric Methods", in Sofia, and challenging the author to present new results. All  colleagues  mentioned in the text have frankly commented upon the manuscript and enhanced its content. Reviewer comments have, no doubt,  much improved the present  version of the ms.

  \end{document}